\documentstyle[12pt,epsfig]{article}  
\begin{document} 
\baselineskip=18pt

\def\la{\mathrel{\mathpalette\fun <}}
\def\ga{\mathrel{\mathpalette\fun >}}
\def\fun#1#2{\lower3.6pt\vbox{\baselineskip0pt\lineskip.9pt
\ialign{$\mathsurround=0pt#1\hfil##\hfil$\crcr#2\crcr\sim\crcr}}} 

\begin{titlepage} 
\begin{center}
{\Large \bf A model of jet quenching in ultrarelativistic 
heavy ion collisions and high-p$_T$ hadron spectra at RHIC} \\

\vspace{4mm}

I.P.~Lokhtin$^a$ and A.M.~Snigirev$^b$ \\
M.V. Lomonosov Moscow State University, D.V. Skobeltsyn Institute of Nuclear 
Physics, \\
119992, Vorobievy Gory, Moscow, Russia 
\end{center}  

\begin{abstract} 
The method to simulate the rescattering and energy loss of hard partons in 
ultrarelativistic heavy ion collisions has been developed. The model 
is a fast Monte-Carlo tool introduced to modify a standard 
PYTHIA jet event. The full heavy ion event is obtained as a superposition of a 
soft hydro-type state and hard multi-jets. The model is applied to analysis of 
the jet quenching pattern at RHIC. 
\end{abstract}

\bigskip

\vspace{100mm}
\noindent
---------------------------------------------\\
$^a$ e-mail: igor@lav01.sinp.msu.ru\\
$^b$ e-mail: snigirev@lav01.sinp.msu.ru 
\end{titlepage}   

\section{Introduction} 

One of the important tools for studying the properties of quark-gluon plasma 
(QGP) in ultrarelativistic heavy ion collisions is the analysis of a QCD 
jet production. 
The medium-induced energy loss of energetic partons, ``jet quenching'', should 
be very different in the cold nuclear matter and QGP, resulting in many 
observable phenomena~\cite{baier_rev}. Recent RHIC data on high-p$_T$ particle 
production~\cite{qm} (the suppression of hadron spectra and azimuthal 
back-to-back two-particle correlations, strong elliptic flow) are in  
agreement with the jet quenching hypothesis~\cite{Wang:2004}. At LHC, 
a new regime of heavy ion physics will be reached at $\sqrt{s_{\rm NN}}=5.5 A$ 
TeV where hard and semi-hard particle production can stand out against the 
underlying soft events. The initial gluon densities in Pb+Pb reactions at LHC 
are expected to be significantly higher than those at RHIC, implying a stronger 
partonic energy loss, observable in various new channels~\cite{Accardi:2003}.

In the most of available Monte-Carlo heavy ion event generators the medium-induced 
partonic rescattering and energy loss are either ignored or implemented 
insufficiently~\cite{mc-lhc}. Thus, in order to analyze RHIC data on 
high-p$_T$ hadron production, test the sensitivity of LHC 
observables to the QGP formation, and study the corresponding experimental 
capabilities of detectors, the development of adequate and fast Monte-Carlo 
models for jet quenching simulation is necessary.

In this paper we present the model of jet quenching and discuss 
its validation basing on RHIC data for high-$p_T$ hadron spectra. In Sect.~2 we
give the physics frameworks of the model. Section 3 describes the event-by-event 
simulation procedure. In Sect.~4 the generalization of the model to the case of 
``full'' heavy ion event (superposition of soft hydro-type state and hard 
multi-jets) is fulfilled. In Sect.~5 the efficiency of the model is 
demonstrated by means of the numerical analysis of hadron spectra at RHIC.  

\section{Physics frameworks of the model} 

The detailed description of physics frameworks of the developed model can be
found in a number of our previous 
papers~\cite{lokhtin98,lokhtin00,lokhtin01,lokhtin02,lokhtin03}. Our approach 
bases on an accumulating energy loss, the gluon radiation being associated 
with each parton scattering in the expanding medium and includes the 
interference effect using the modified radiation spectrum $dE/dl$ as a function 
of decreasing temperature $T$. The basic kinetic integral equation for the 
energy loss $\Delta E$ as a function of initial energy $E$ and path length $L$ 
has the form 
\begin{eqnarray} 
\label{elos_kin}
\Delta E (L,E) = \int\limits_0^Ldl\frac{dP(l)}{dl}
\lambda(l)\frac{dE(l,E)}{dl} \, , ~~~~ 
\frac{dP(l)}{dl} = \frac{1}{\lambda(l)}\exp{\left( -l/\lambda(l)\right) }
\, ,  
\end{eqnarray} 
where $l$ is the current transverse coordinate of a parton, $dP/dl$ is the 
scattering probability density, $dE/dl$ is the energy loss per unit length, 
$\lambda = 1/(\sigma \rho)$ is in-medium mean free path, $\rho \propto T^3$ is 
the medium density at the temperature $T$, $\sigma$ is the integral cross 
section for parton interaction in the medium. 

The collisional energy loss due to elastic scattering with high-momentum 
transfer has been originally estimated by Bjorken in~\cite{bjork82}, and 
recalculated later in~\cite{mrow91} taking also into account the  low-momentum 
transfer loss resulting mainly from the interactions with plasma collective 
modes. Since the latter process does not contribute much to the total 
collisional loss in comparison with high-momentum scattering (due to absence of 
large factor $\sim \ln{(E / \mu_D)}$ where $\mu_D$ is the Debye screening mass) 
and, in numerical estimates it can be effectively ``absorbed'' by means of 
redefinition of minimum momentum transfer $t_{\rm min} \sim \mu_D^2$ , we used 
the collisional part associated with high-momentum transfer only~\cite{lokhtin00},   
\begin{equation} 
\label{col} 
\frac{dE}{dl}^{col} = \frac{1}{4T \lambda \sigma} 
\int\limits_{\displaystyle \mu^2_D}^
{\displaystyle t_{\rm max}}dt\frac{d\sigma }{dt}t ~,
\end{equation} 
and the dominant contribution to the differential cross section 
\begin{equation} 
\label{sigt} 
\frac{d\sigma }{dt} \cong C \frac{2\pi\alpha_s^2(t)}{t^2} 
\frac{E^2}{E^2-m_p^2}~,~~~~
\alpha_s = \frac{12\pi}{(33-2N_f)\ln{(t/\Lambda_{QCD}^2)}} \>
\end{equation} 
for scattering of a hard parton with energy $E$ and mass $m_p$ off the 
``thermal'' parton with 
energy (or effective mass) $m_0 \sim 3T \ll E$. Here $C = 9/4, 1, 4/9$ for $gg$, $gq$ and 
$qq$ scatterings respectively, $\alpha_s$ is the QCD running coupling constant 
for $N_f$ active quark flavors, and $\Lambda_{QCD}$ is the QCD scale parameter 
which is of the order of the critical temperature,  $\Lambda_{QCD}\simeq T_c 
\simeq 200$ MeV. The integrated cross section $\sigma$ is regularized by the 
Debye screening mass squared $\mu_D^2 (T) \simeq 4\pi \alpha _s T^2(1+N_f/6)$. 
The maximum momentum transfer $t_{\rm max}=[ s-(m_p+m_0)^2] [ s-(m_p-m_0)^2 ] / 
s$ where $s=2m_0E+m_0^2+m_p^2$.

There are several calculations of the inclusive energy distribution of 
medium-induced gluon radiation using Feyman multiple scattering diagrams. The 
relation between these approaches and their basic parameters has been discussed 
in detail in the recent writeup of the working group ``Jet Physics'' for the 
CERN Yellow Report~\cite{Accardi:2003}. We restrict ourselves to using 
BDMS formalism~\cite{baier}. In the BDMS frameworks, the strength of multiple 
scattering is characterized by the transport coefficient 
$\hat{q}=\mu_D^2/\lambda_g $ ($\lambda_g$ is the gluon mean free path), which is
related to the elastic scattering cross section $\sigma$ (\ref{sigt}). In our
simulations this strength is in fact regulated mainly by the initial QGP 
temperature $T_0$. Then the energy spectrum of coherent medium-induced gluon 
radiation and the corresponding dominant part of radiative energy loss of
massless parton have the form~\cite{baier}: 
\begin{eqnarray} 
\label{radiat} 
\frac{dE}{dl}^{rad} = \frac{2 \alpha_s (\mu_D^2) C_R}{\pi L}
\int\limits_{\omega_{\min}}^E  
d \omega \left[ 1 - y + \frac{y^2}{2} \right] 
\>\ln{\left| \cos{(\omega_1\tau_1)} \right|} 
\>, \\  
\omega_1 = \sqrt{i \left( 1 - y + \frac{C_R}{3}y^2 \right)   
\bar{\kappa}\ln{\frac{16}{\bar{\kappa}}}}
\quad \mbox{with}\quad 
\bar{\kappa} = \frac{\mu_D^2\lambda_g  }{\omega(1-y)} ~, 
\end{eqnarray} 
where $\tau_1=L/(2\lambda_g)$, $y=\omega/E$ is the fraction of the hard parton 
energy carried away by the radiated gluon, and $C_R = 4/3$ is the quark color 
factor. A similar expression for the gluon jet can be obtained by setting  
$C_R=3$ and proper by changing the factor in the square brackets in (\ref{radiat}), 
see ref.~\cite{baier}. The integration (\ref{radiat}) is carried out over all 
energies from $\omega_{\min}=E_{\rm LPM}=\mu_D^2\lambda_g$, the minimum radiated 
gluon energy in the coherent LPM regime, up to initial parton energy $E$. 
Note that we do not consider here possible effects of  
double parton scattering~\cite{Wang:2001,Vitev:2004} and thermal 
gluon absorption~\cite{Wang:2001-2}, which can be included in the model in   
the future. 

The simplest generalization of the formula for a heavy quark of mass $m_q$ can 
be done by using the ``dead-cone'' approximation~\cite{dc}:  
\begin{equation}
\label{radmass} 
\frac{dE}{dld\omega }| _{m_q \ne 0} =  \frac{1}{(1+(\beta \omega )^{3/2})^2}
\frac{dE}{dld\omega }| _{m_q=0}, ~~~ \beta =\left( \frac{\lambda}{\mu_D^2}\right) ^{1/3}
\left( \frac{m_q}{E}\right) ^{4/3}~. 
\end{equation}
One should mention the more recent developments on heavy quark energy loss 
calculations available in the literature~\cite{djor,armesto,zhang}, which can be 
also considered as further model improvements.

The medium is treated as a boost-invariant longitudinally expanding quark-gluon 
fluid, and partons as being produced on a hyper-surface of equal proper times 
$\tau$~\cite{bjorken}. In order to simplify numerical calculations we omit here 
the transverse expansion and viscosity of the fluid using the well-known scaling 
solution obtained by Bjorken~\cite{bjorken} for a temperature and density of QGP 
at $T > T_c \simeq 200$ MeV:
\begin{equation}
\varepsilon(\tau) \tau^{4/3} = \varepsilon_0 \tau_0^{4/3},~~~~
T(\tau) \tau^{1/3} = T_0 \tau_0^{1/3},~~~~ \rho(\tau) \tau = \rho_0 \tau_0 .
\end{equation}
The internal model parameters are the initial conditions for the QGP  
formation expected for central Au+Au (Pb+Pb) collisions at RHIC (LHC): 
$\tau_0$, $T_0$ and $N_f$.  
For non-central collisions and for other beam atomic numbers we suggest 
the proportionality of the initial energy density $\varepsilon _0$ to the ratio 
of nuclear overlap function and effective transverse area of nuclear 
overlapping~\cite{lokhtin00}.

Note that using other scenarios of QGP space-time evolution for the Monte-Carlo 
implementation of the model is also envisaged. In fact, the influence of 
the transverse flow, as well as that of the mixed phase at $T = T_c$, on the 
intensity of jet rescattering (which is a strongly increasing function of $T$) 
has been found to be inessential for high initial temperatures $T_0 \gg T_c$. 
On the contrary, the presence of QGP viscosity slows down the cooling rate, 
that implies a jet parton spending more time in the hottest regions of the 
medium. As a result the rescattering intensity increases, i.e., in fact an 
effective temperature of the medium gets higher as compared with the perfect QGP 
case. We also do not take into account here the probability of jet rescattering 
in nuclear matter, because the intensity of this process and corresponding 
contribution to the total energy loss are not significant due to much smaller 
energy density in a ``cold'' nucleus.

Another important element of the model is the angular spectrum of in-medium gluon 
radiation. Since the detailed calculation of the angular spectrum of emitted gluons 
is rather sophisticated and 
model-dependent~\cite{lokhtin98,baier,Vitev:2004,Zakharov:1999,urs,vitev}, 
the simple parameterization of gluon angular distribution over the emission 
angle $\theta$ was used:
\begin{equation} 
\label{sar} 
\frac{dN^g}{d\theta}\propto \sin{\theta} \exp{\left( -\frac{(\theta-\theta
_0)^2}{2\theta_0^2}\right) }~, 
\end{equation}
where $\theta_0 \sim 5^0$ is the typical angle of the coherent gluon radiation 
as estimated in~\cite{lokhtin98}. Other parameterizations are also envisaged. 

\section{Simulation procedure} 

The model has been constructed as the fast Monte-Carlo event generator PYQUEN 
(PYthia QUENched), and the corresponding Fortran routine PYQUEN is available 
via Internet~\cite{pyquen}. The routine is implemented as a modification of 
the standard PYTHIA$\_$6.2.* jet event~\cite{pythia}. 

The following event-by-event Monte-Carlo simulation procedure is applied. \\
$\bullet$ Generation of the initial parton spectra with PYTHIA (fragmentation 
{\em off}). \\ 
$\bullet$ Generation of the jet production vertex at the impact parameter 
$b$ according to the distribution
\begin{equation} 
\frac{dN^{\rm jet}}{d\psi dr} (b) = \frac{T_A(r_1) T_A(r_2)}{T_{AA}(b)}, ~~~~ 
T_{AA}(b)=\int\limits_0^{2\pi} d \psi \int\limits_0^{r_{max}}r dr T_A(r_1) 
T_A(r_2) ~,
\end{equation} 
where $r_{1,2} (b,r,\psi)$ are the distances between the nucleus centers and 
the jet production vertex $V(r\cos{\psi}, r\sin{\psi})$; $r_{max} (b, \psi) 
\le R_A$ is the maximum possible transverse distance $r$ from the nuclear 
collision axis to $V$; $R_A$ is the radius of the nucleus $A$; 
$T_A({\bf r}) = A \int \rho_A({\bf r},z)dz$ is the nuclear thickness function 
with nucleon density distribution $\rho_A({\bf r},z)$; $T_{AA}(b)$ is the nuclear overlap function  
(see ref.~\cite{lokhtin00} for detailed nuclear geometry explanations). \\ 
$\bullet$ Calculation of scattering cross section $\sigma = \int dt~d\sigma/dt$ 
(\ref{sigt}). \\
$\bullet$ Generation of the displacement between $i$-th and $(i+1)$-th
scatterings, $l_i = (\tau_{i+1} - \tau_i)$:  
\begin{equation} 
\frac{dP}{dl_i} = \lambda^{-1}(\tau_{i+1}) \exp{(-\int\limits_0^{l_i}
\lambda^{-1} (\tau_i + s)ds)} ~,~~ \lambda^{-1}(\tau ) =\sigma (\tau ) \rho 
(\tau )~,
\end{equation}  
and calculation of the corresponding transverse distance, $l_i p_T/E$. \\
$\bullet$ Reducing the parton energy by collisional and radiative loss per 
each $i$-th scattering:
\begin{equation} 
\Delta E_{{\rm tot},i} = \Delta E_{{\rm col},i} + \Delta E_{{\rm rad},i}~,
\end{equation} 
where the collisional part is calculated in the high-momentum transfer 
approximation (\ref{sigt}), 
\begin{equation} 
\Delta E_{{\rm col},i} = \frac{t_i}{2 m_0}~,
\end{equation} 
and the energy of a radiated gluon, $\omega _i=\Delta E_{{\rm rad},i}$, is 
generated according to (\ref{radiat}) and (\ref{radmass}) :
\begin{eqnarray}
\frac{dI}{d\omega }| _{m_q=0} = \frac{2 \alpha_s (\mu_D^2) 
\lambda C_R}{\pi L \omega } \left[ 1 - y + \frac{y^2}{2} \right] 
\>\ln{\left| \cos{(\omega_1\tau_1)} \right|}~, 
~~\frac{dI}{d\omega }| _{m_q \ne 0} =  \frac{1}{(1+(\beta \omega )^{3/2})^2}
\frac{dI}{d\omega }| _{m_q=0}~.  
\end{eqnarray}
$\bullet$ Calculation of the parton transverse momentum kick due to elastic 
scattering $i$:
\begin{equation} 
\Delta k_{t,i}^2 =(E-\frac{t_i}{2m_{0i}})^2-(p-\frac{E}{p}\frac{t_i}{2m_{0i}}-
\frac{t_i}{2p})^2-m_p^2 .
\end{equation} 
$\bullet$ Formation of the additional (in-medium emitted) gluon with the energy 
$\omega _i$ and the emission angle $\theta _i$ relative to the parent parton   
determined according to the parameterization (\ref{sar}). \\
$\bullet$ Halting the rescattering if {\em (1)} the parton escapes the 
dense zone, or {\em (2)} QGP cools down to $T_c=200$ MeV, or {\em (3)} the 
parton loses so much energy that its $p_T (\tau)$ drops below $2T (\tau)$. \\
$\bullet$ At the end of each event, adding new (in-medium emitted) gluons to 
the PYTHIA parton list and rearrangement of partons to update string 
formation. \\
$\bullet$ Formation of the final state particles by PYTHIA 
(fragmentation {\em on}).  

\section{Extension of the model to simulate full heavy ion event}  

The full heavy ion event is simulated as a superposition of soft hydro-type 
state and hard multi-jets. The simple 
approximation\cite{Kruglov:1997,Lokhtin:2002} of hadronic liquid at  
``freeze-out'' stage has been used to treat soft part of the event giving the 
final hadron spectrum as a superposition of a thermal distribution and 
a collective flow~\cite{Heinz:1993,Muroya:1995,Kolb:2000}. \\ 
1. The 4-momentum $p^*_{\mu}$ of a hadron of mass $m$ was generated at random 
in the rest frame of a liquid element in accordance with the isotropic 
Boltzmann distribution 
\begin{eqnarray}
f(E^*) \propto E^*\sqrt{E^{*2}-m^2}\exp{(-E^*/T_f)}, ~~~~-1 < \cos{\theta^*} < 
1, ~~~~ 0 < \phi^* < 2\pi ~~, 
\end{eqnarray}  
where $E^*=\sqrt{p^{*2}+m^2}$ is the energy of the hadron, and the polar angle 
$\theta^*$ and the azimuthal angle $\phi^*$ specify the direction of its motion
in the rest frame of the liquid element. \\  
2. The spatial position of a liquid element and its local 4-velocity 
$u_{\mu}$ were generated at random in
accordance with phase space and the character of motion of the fluid:  
\begin{eqnarray}
\label{space}
&     & f(r) = 2r/R_f^2 ~(0 < r < R_f), ~~~~f(\eta )\propto 
\exp\left[ {-(\eta -Y_L^{\rm max})^2/2(Y_L^{\rm max})^2}\right] , ~~~~
0 < \Phi < 2\pi, \nonumber \\ 
&     & u_r = \sinh{Y_T^{\rm max}}\frac{r}{\sqrt{R_f(b)R_f(b=0)}}, 
~~~~u_t=\sqrt{1+u_r^2}\cosh{\eta}, ~~~~ 
u_z=\sqrt{1+u_r^2}\sinh{\eta} ~~,    
\end{eqnarray}  
where $R_f$ is the final transverse radius of the system in a given direction.  
Freeze-out parameters of the model are kinetic freeze-out 
temperature $T_f$ and maximum longitudinal, $Y_L^{\rm max}$, and transverse, 
$Y_T^{\rm max}$, collective flow rapidities. \\   
3. Further, boost of the hadron 4-momentum in the 
c.m. frame of the event was calculated: 
\begin{eqnarray}    
p_x & = & p^*\sin{\theta^*}\cos{\phi^*} + u_r\cos{\Phi}\left[ E^* + \frac{(u^i
p^{*i})}{u_t + 1}\right] \nonumber \\ 
p_y & = & p^*\sin{\theta^*}\sin{\phi^*} + u_r\sin{\Phi}\left[ E^* + \frac{(u^i
p^{*i})}{u_t + 1}\right] \nonumber \\ 
p_z & = & p^*\cos{\theta^*} + u_z\left[ E^* + \frac{(u^i 
p^{*i})}{u_t + 1}\right] \nonumber \\ 
E & = & E^*u_t + (u^i p^{*i}),    
\end{eqnarray} 
where 
\begin{eqnarray}
(u^i p^{*i}) & = & u_rp^*\sin{\theta^*}\cos{(\Phi-\phi^*)} + u_z p^*
\cos{\theta^*} ~.     
\end{eqnarray} 
Anisotropic flow is introduced here under simple assumption that the spatial 
ellipticity of ``freeze-out'' region, 
$\epsilon =\left< y^2 - x^2 \right> / \left< y^2 + x^2 \right> $,
is directly related to the ellipticity of the system formed in the region of the
initial overlap of nuclei, 
$\epsilon _0 = b/2R_A$.  
This ``scaling'' enables one to avoid introducing additional parameters
and, at the same time, leads to an azimuthal anisotropy of generated particles 
due to dependence of transverse radius $R_f (b)$ on the angle  
$\Phi$~\cite{Lokhtin:2002}: 
\begin{equation}
\label{R_f} 
R_f(b) = R_{f}(b=0)~{\rm min}\{ \sqrt{1 - \epsilon^2_0~ \sin^2 \Phi} + \epsilon_0~ 
\cos \Phi , ~~ \sqrt{1 - \epsilon^2_0~ \sin^2 \Phi} - \epsilon_0~ 
\cos \Phi \} . 
\end{equation} 
Obtained in such a way azimuthal distribution of particles is described well by 
the elliptic form for the domain of reasonable impact parameter values.  

The mean total particle multiplicity in central Au+Au (Pb+Pb) collisions at 
RHIC (LHC) is the input parameter of the model (instead of $R_f(b=0)$ we put  
$R_A$ here for simplicity), the total multiplicity for other 
centralities and atomic numbers being assumed to be proportional to the number of
nucleons-participants. We also set the Poisson multiplicity distribution and 
the following particle ratios:  
$$\pi^{\pm}:K^{\pm}:p^{\pm} = 24:6:1,~~~~~\pi^{\pm}:\pi^0=2:1,
~~~~~K^{\pm}:K^0=1:1,~~~~~p:n=1:1~~.$$   

The hard part of the event includes PYTHIA/PYQUEN hadronic jets generated  
according to the binomial distribution. The mean number of jets produced in 
AA events at a given $b$ is proportional to the number of binary 
nucleon-nucleon sub-collisions and determined as 
\begin{equation}
\label{numjets}
\overline{N_{AA}^{{\rm jet}}} (b,\sqrt{s}) =  
T_{AA}(b) \int\limits_{p_T^{\rm min}} dp_T^2 \int dy  
\frac{d\sigma_{pp}^{\rm hard}(p_T,~\sqrt{s})}{dp_T^2dy}~, 
\end{equation} 
where $d\sigma_{pp}^{\rm hard}(p_T, \sqrt{s})/dp_T^2dy$ is the cross section of 
corresponding hard process in $pp$ collisions (at the same c.m.s. energy, 
$\sqrt{s}$, of colliding beams) with the minimum transverse momentum transfer 
$p_T^{\rm min}$. The latter is another input parameter of the model. In the 
frameworks of our approximation, partons produced in (semi)hard
processes with the momentum transfer less than $p_T^{\rm min}$ are considered 
as being ``thermalized'', so their hadronization products are included in a   
soft part of the event ``automatically''. 

Note that we can expect some adequate results only for central and semi-central 
collisions, but not for very peripheral collisions ($b \sim 2 R_A$) where
the hydro-type description is not applicable. Besides, the very forward 
rapidity region (where other dynamical effects can be important) is beyond  
our treatment here. 

The extended in such a way jet quenching model has been constructed as the 
fast Monte-Carlo event generator, and the corresponding Fortran code is also 
available via Internet~\cite{hydjet}.

Let us remind in the end of this section, that ideologically our approximation
is similar to the model of Hirano and Nara~\cite{hirano}. The difference is 
that we concentrate here on the detailed simulation of the parton multiple 
scattering in a QCD-medium (the scattering-by-scattering generation of parton 
path length and energy loss in an expanding QGP, taking into account the 
collisional loss, Lund string fragmentation model both for hard partons and 
in-medium emitted gluons, etc.), while the treatment of the hydrodynamic part 
in~\cite{hirano} is much more detailed than in our simple (and therefore fast) 
simulation procedure.

\section{Validation of the model at RHIC, $\sqrt{s}=200~A$ GeV}

In order to demonstrate the efficiency of the model, the jet quenching pattern 
in Au+Au collisions at RHIC was considered. The comparison of calculated and
experimentally measured pseudorapidity $\eta$ and transverse momentum $p_T$ 
spectra of hadrons together with their dependence of  
event centrality  allows the optimization of the model and
specification of main model parameters. 

The PHOBOS data on $\eta$-spectra of charged hadrons~\cite{phobos} have been 
analyzed to fix the particle density in the mid-rapidity region and the maximum 
longitudinal flow rapidity, $Y_L^{\rm{max}}=3.5$. For the calculation of 
(multi)jet production cross section, we used the factor $K=2$ taking into 
account higher order corrections of perturbative QCD. The rest of the model 
parameters 
have been obtained by fitting PHENIX data on $p_T$-spectra of neutral 
pions~\cite{phenix}: the kinetic freeze-out temperature $T_f=100$ MeV, maximum 
transverse flow rapidity $Y_T^{\rm{max}}=1.25$ and minimum transverse momentum 
transfer of ``non-thermalized'' hard process  $p_T^{\rm min}=2.8$ GeV/$c$. 
It was found that the nuclear modification of the hardest domain of 
$p_T$-spectrum ($\ga 5$ GeV/$c$) is determined in our case only by the 
intensity of the medium-induced parton rescattering. This fact allows us to 
extract from the data initial conditions of the QGP formation independently on 
other input parameters: the initial temperature $T_0=500$ MeV, the formation time 
$\tau_ 0=0.4$ fm/$c$ and the number of active quark flavours $N_f=2$. We will see 
below, that setting model parameters as it was described above, makes it 
possible to reproduce the main features of jet quenching pattern at RHIC: 
the $p_T$--dependence of the nuclear modification factor $R_{AA}$ and 
two-particle azimuthal correlation function $C(\Delta \varphi)$.   

Figure 1 shows $\eta$-distribution of charged hadrons in Au+Au collisions for 
different centrality sets. The good fit of PHOBOS data~\cite{phobos}
is achieved excepting very forward rapidities. The  
$p_T$-distributions of $\pi ^0$-mesons obtained at PHENIX~\cite{phenix} is also
well reproduced by our calculations, even for relatively peripheral collisions 
(Figure 2). 

Figure 3 shows the nuclear modification factor $R_{AA}$ for neutral pions, 
which is defined as a ratio of particle yields in $AA$ and $pp$ 
collisions normalized on the number of binary nucleon-nucleon sub-collisions: 
\begin{equation} 
\label{raa} 
R_{AA}=\frac{d\sigma^{\pi^0} _{AA}/dp_T}{T_{AA}(b) 
\sigma _{\rm in}d\sigma ^{\pi^0}_{pp}/dp_T}~,
\end{equation} 
where $\sigma _{\rm in}=42$ mb is the inelastic non-diffractive $pp$ cross 
section at $\sqrt{s}=200$ GeV. In the absence of medium-induced effects in the 
mid-rapidity region it should be $R_{AA}=1$ for high enough 
$p_T(\ga 2$ GeV/$c$). Such value of $R_{AA}sim 1$ 
has been observed so for d+Au and peripheral Au+Au collisions, but not for 
central and semi-central Au+Au events, where $R_{AA} <1$ up to maximum 
measured transverse momenta $p_T \sim 10$ GeV/$c$. One can see from Figure 3 
that our model calculations reproduce $p_T$-- and centrality dependences of
$R_{AA}$~\cite{phenix} quite well.  

Another important tool to verify jet quenching is two-particle azimuthal 
correlation function $C(\Delta \varphi)$ -- the distribution over an azimuthal 
angle of high-$p_T$ hadrons in the event with $2$ GeV/$c<p_T<p_T^{\rm trig}$  
relative to that for the hardest ``trigger'' particle with $p_T^{\rm trig}>4$ 
GeV/$c$. 
Figure 4 presents $C(\Delta \varphi)$ in $pp$ and in central Au+Au collisions 
(data from STAR~\cite{star}). Clear peaks in $pp$ collisions at $\Delta \varphi 
=0$ and $\Delta \varphi =\pi$ indicate a typical dijet event topology. Note 
that almost the same pattern has been observed in d+Au and peripheral Au+Au 
collisions. However, for most central Au+Au collisions the peak near $\pi$ 
disappears. It can be interpreted as the observation of monojet events due to the
``absorption'' of one of the jets in a dense medium. Such event configuration
corresponds to the situation when the dijet production vertex is close to the 
surface of the nuclear overlap region: then one partonic jet can escape  
the medium almost without re-interactions and then go to detectors, while second 
jet loses most of its initial energy due to a large number of 
rescatterings and therefore becomes unobservable~\cite{Plumer:1995}. 
Figure 4 demonstrates that measured suppression of azimuthal back-to-back
correlations is well reproduced by our model (the same procedure of 
uncorrelated background subtraction as in~\cite{star} was applied).  

We leave beyond the scope of this paper the analysis of such important RHIC 
observables as the azimuthal anisotropy and particle ratios. Since these 
observables are very sensitive to the soft physics, in order to study them  
a more careful treatment of low-$p_T$ particle production than our simple 
approach is needed (the detailed description of space-time structure of 
freeze-out region, resonance decays, etc.) For example, our model can reproduce 
experimentally measured~\cite{phenix-v2,star-v2} $p_T$-dependence of the 
coefficient of azimuthal anisotropy $v_2$ (the second harmonic of Fourier 
decomposition of particle azimuthal distribution) qualitatively, giving rapid 
hydrodynamical growth up to $p_T \sim 3$ GeV/$c$ with the subsequent 
saturation. However, the model calculations significantly underestimate the 
data at $p_T<2$ GeV/$c$. A solution of baryon-to-meson ratio 
``puzzle''~\cite{phenix-pr,star-pr} is also beyond our consideration here. 
Further development of our model with special emphasis on the more detailed 
description of low-$p_T$ particle production is planed for the future. 

\section{Conclusions} 
The model of jet quenching in ultrarelativistic heavy ion collisions has been 
developed. It includes the generation of the hard parton production vertex
according to the realistic nuclear geometry, rescattering-by-rescattering 
simulation of the parton path length in a dense matter, radiative and 
collisional energy loss per rescattering, final hadronization with the Lund 
string fragmentation model for hard partons and in-medium emitted gluons. 
The model is the fast Monte-Carlo tool implemented to modify 
a standard PYTHIA jet event. The model has been generalized to the case of 
the ``full'' heavy ion event (the superposition of soft, hydro-type state and 
hard multi-jets) using a simple and fast simulation procedure for soft 
particle production.  

The efficiency of the model is demonstrated basing on the numerical analysis of 
high-$p_T$ hadron production in Au+Au collisions at RHIC. The good fit of
experimental data on $\eta$-- and $p_T$-- spectra of hadrons for different event 
centralities is achieved. The model is capable of reproducing main features of 
the jet quenching pattern at RHIC: the $p_T$ dependence of the nuclear 
modification factor $R_{AA}$, and the suppression of azimuthal back-to-back 
correlations. The further development of the model focusing on a more detailed 
description of low-$p_T$ particle production is planed for the future. 

\bigskip

{\it Acknowledgments}.  Discussions with A.I.~Demianov, Yu.L.~Dokshitzer, 
A.~Morsch, 
S.V.~Petru\-shan\-ko, C.~Ro\-land, L.I.~Sarycheva, J.~Schukraft, C.Yu.~Teplov,
I.N.~Vardanyan, I.~Vitev, B.~Wyslouch, B.G.~Zakharov and G.M.~Zinovjev and 
are gratefully acknowledged. This work is supported by grant 
N 04-02-16333 of Russian Foundation for Basic Research.

\begin{figure}[hbtp] 
\begin{center} 
\makebox{\epsfig{file=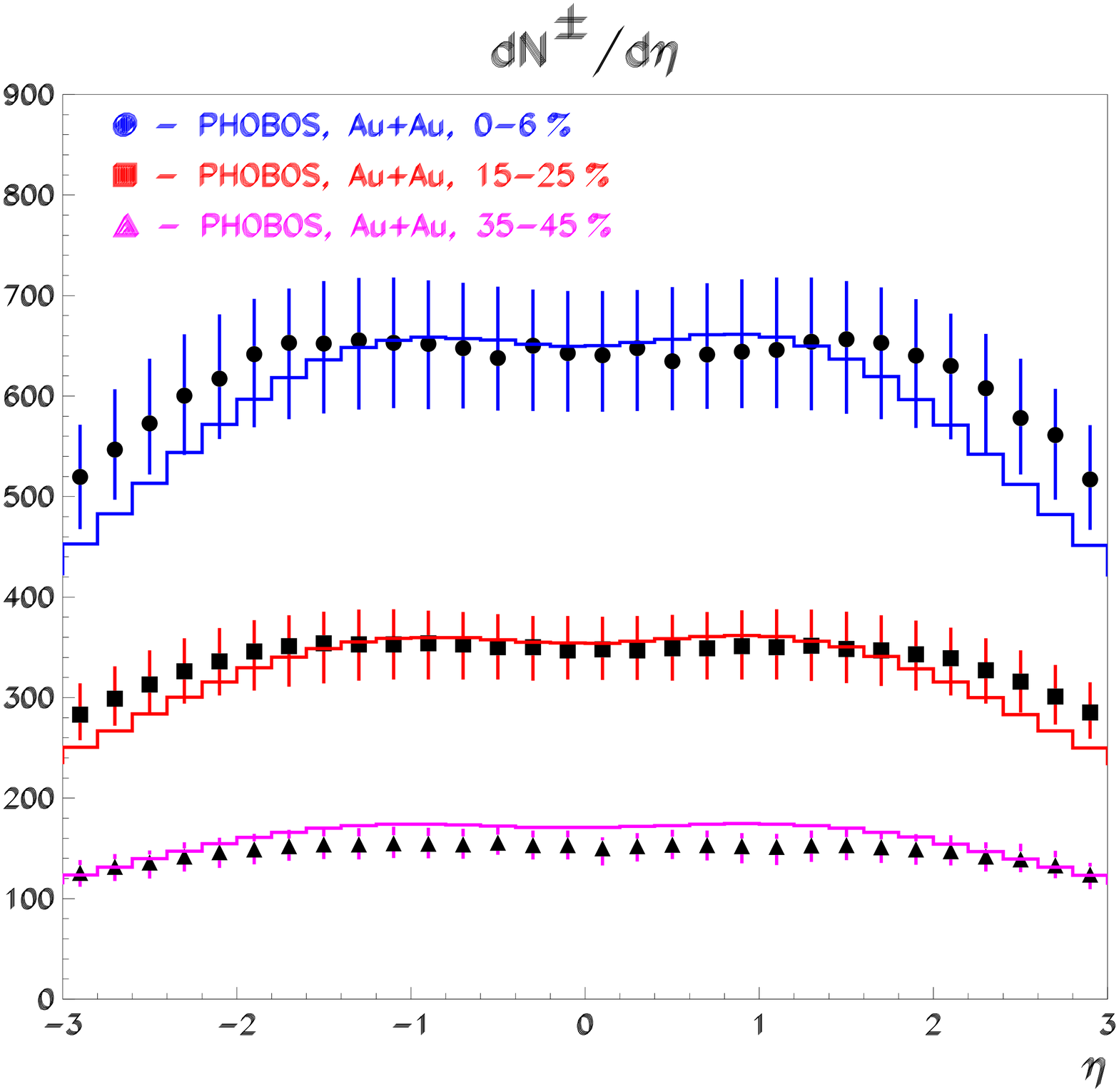, height=170mm}}   
\vskip 1cm 
\caption{\small The pseudorapidity distribution of charged hadrons in Au+Au 
collisions for three centrality sets. The points are PHOBOS data~\cite{phobos}, 
histograms are the model calculations.} 
\end{center}
\end{figure}

\begin{figure}[hbtp] 
\begin{center} 
\makebox{\epsfig{file=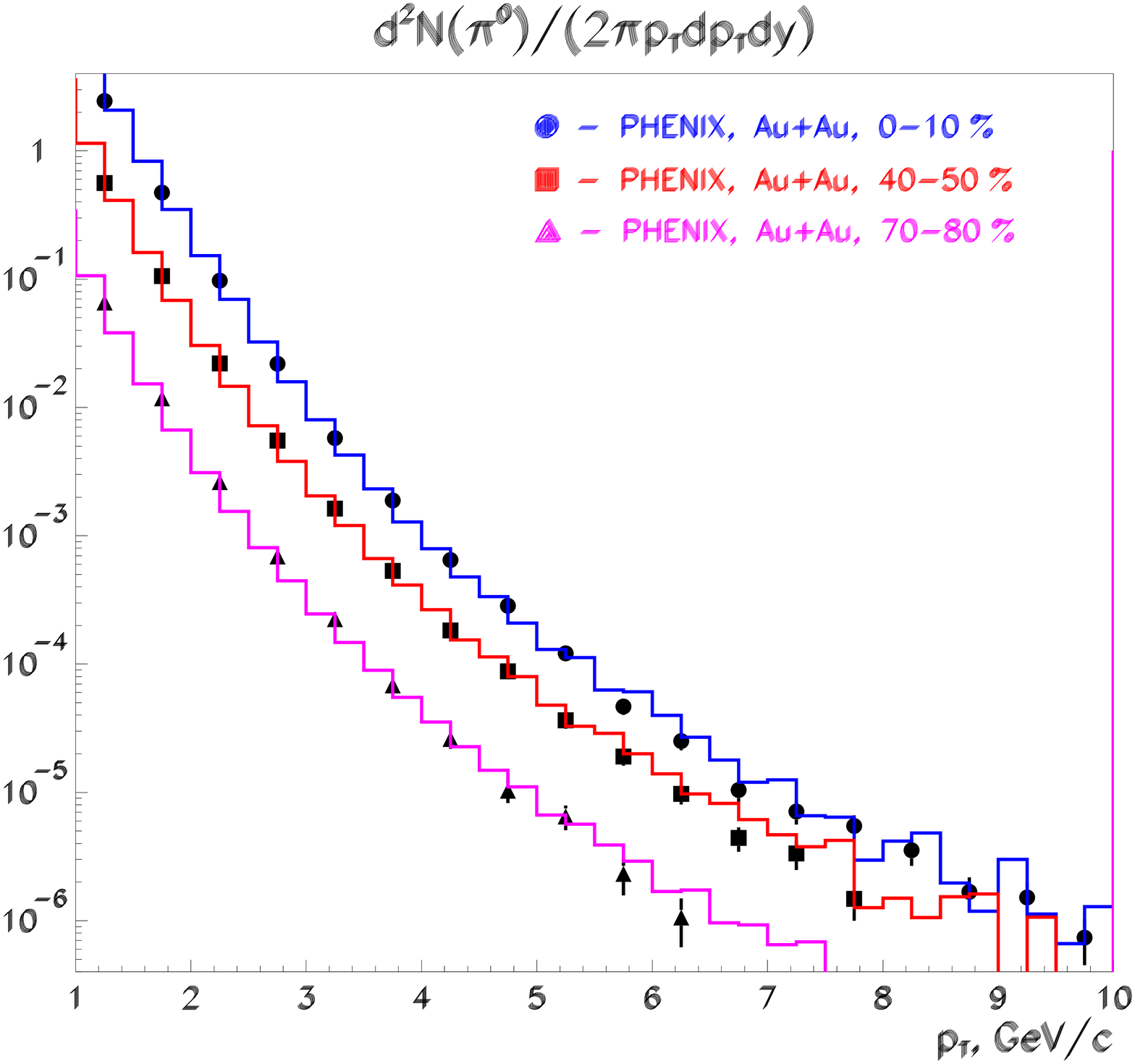, height=170mm}}   
\vskip 1cm 
\caption{\small The transverse momentum distribution of neutral pions in Au+Au 
collisions for three centrality sets. The points are PHENIX data~\cite{phenix}, 
histograms are the model calculations.} 
\end{center}
\end{figure}

\begin{figure}[hbtp] 
\begin{center} 
\makebox{\epsfig{file=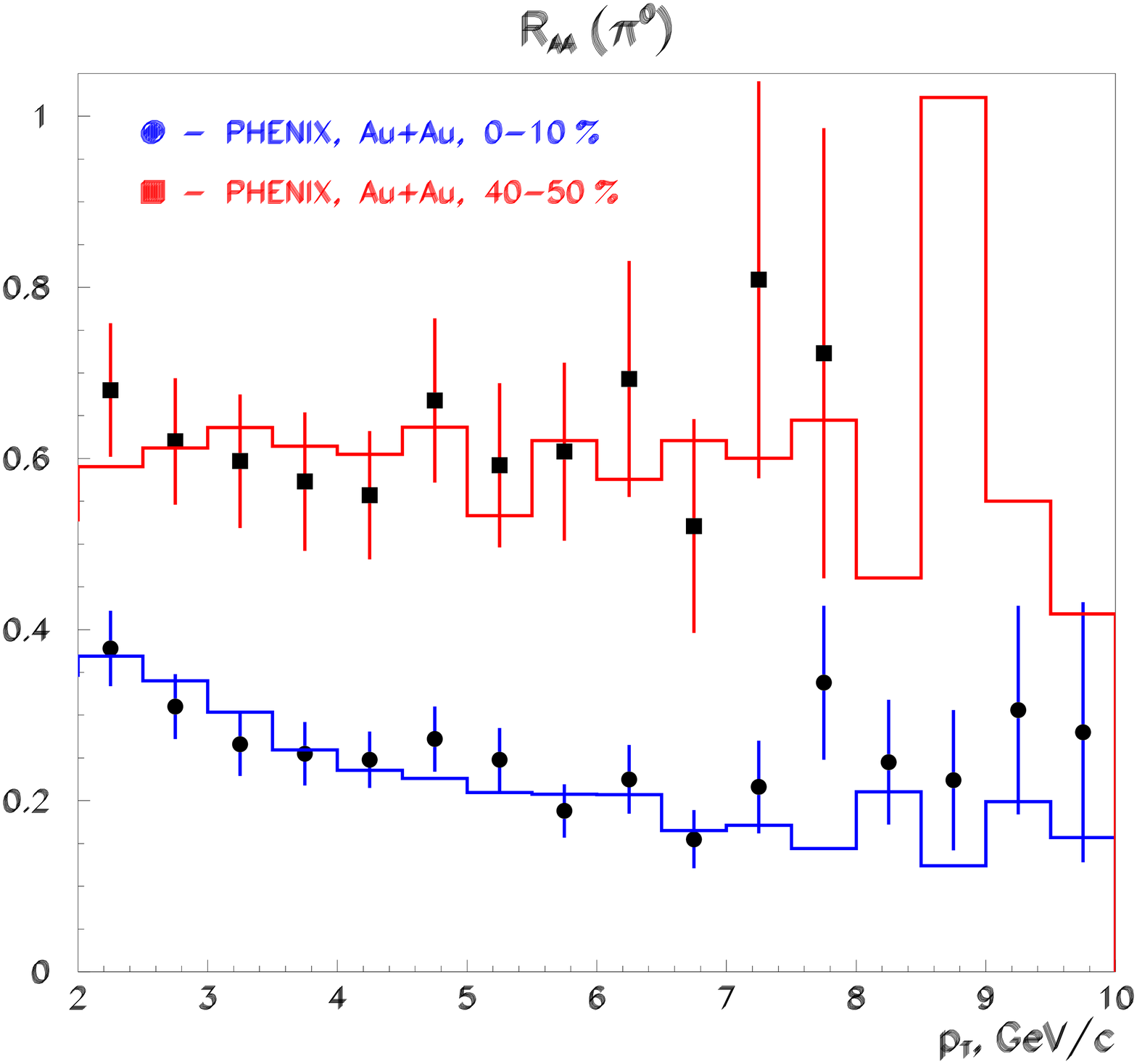, height=170mm}}   
\vskip 1cm 
\caption{\small The nuclear modification factor $R_{AA}$ (\ref{raa}) for 
neutral pions in Au+Au collisions for two centrality sets. The points are  
PHENIX data~\cite{phenix}, histograms are the model calculations. } 
\end{center}
\end{figure}

\begin{figure}[hbtp] 
\begin{center} 
\makebox{\epsfig{file=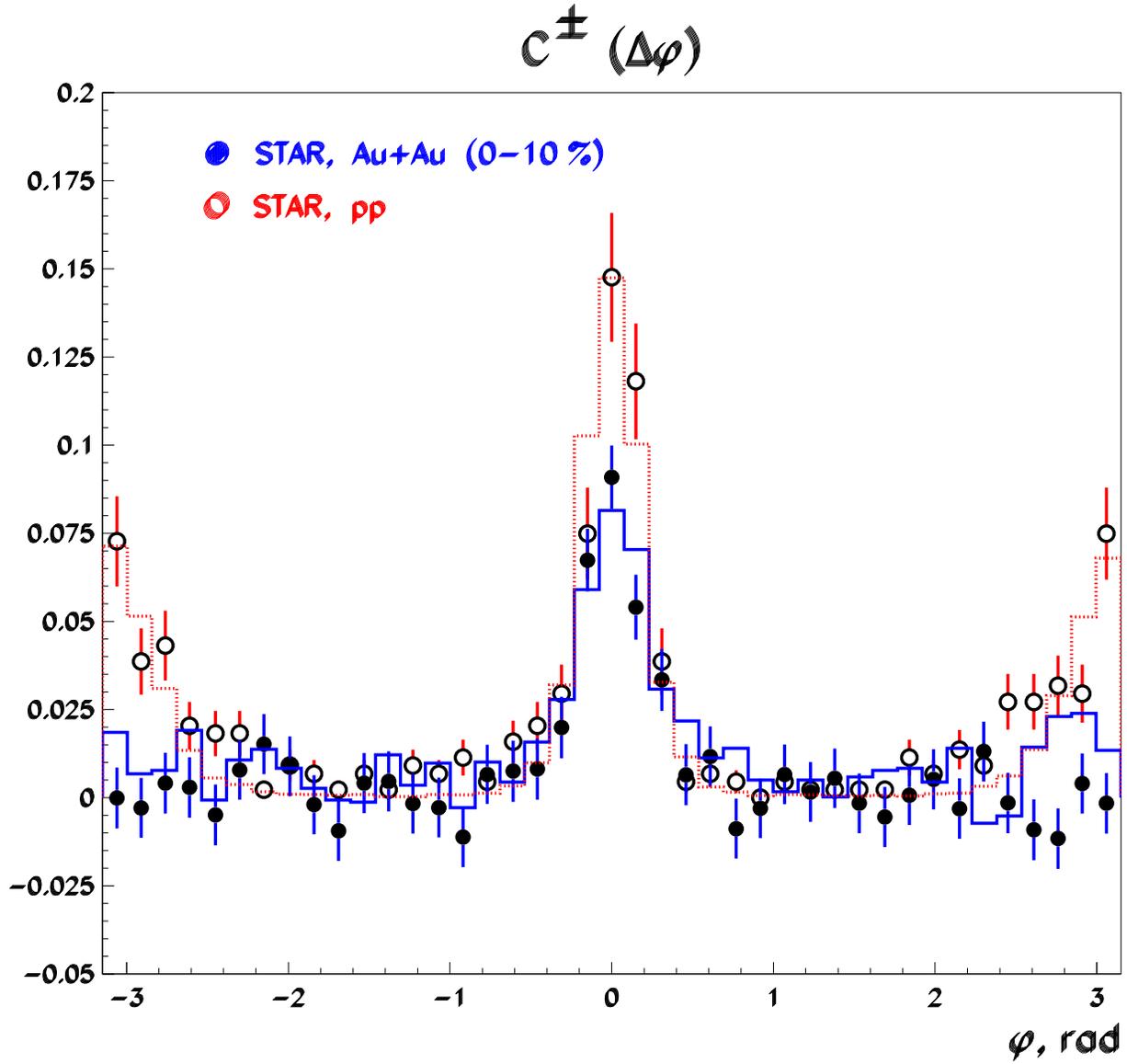, height=170mm}}   
\vskip 1cm 
\caption{\small The azimuthal two-particle correlation function for $pp$ and for 
central Au+Au collisions. The points are STAR data~\cite{phenix}, dashed and 
solid histograms are the model calculations for $pp$ and Au+Au events 
respectively.} 
\end{center}
\end{figure}

\end{document}